\documentclass[12pt]{iopart} 

\expandafter\let\csname equation*\endcsname\relax 
\expandafter\let\csname endequation*\endcsname\relax 

\usepackage{amsfonts}
\usepackage{graphicx}
\usepackage{amsmath}
\usepackage{amssymb}
\usepackage{color}

\def\beq{\begin{equation}}
\def\eeq{\end{equation}}
\def\bea{\begin{eqnarray}}
\def\eea{\end{eqnarray}}
\def\ben{\begin{enumerate}}
\def\een{\end{enumerate}}

\begin{document}

\title[Exact Solutions to Force-Free Electrodynamics]{Exact Solutions to Force-Free Electrodynamics in Black Hole Backgrounds}

\author{T. Daniel Brennan}

\author{Samuel E. Gralla}

\author{Ted Jacobson}

\address{Maryland Center for Fundamental Physics \& Joint Space-Science Institute, Department of Physics, University of Maryland, College Park, MD 20742, USA}

\begin{abstract}

A shared property of several of the known exact solutions to the equations of force-free electrodynamics is that their charge-current four-vector is \textit{null}.  We examine the general properties of null-current solutions and then focus on the principal congruences of the Kerr black hole spacetime.  We obtain a large class of exact solutions, which are in general time-dependent and non-axisymmetric.   These solutions include waves that, surprisingly, propagate without scattering on the curvature of the black hole's background.  They may be understood as generalizations to Robinson's solutions to vacuum electrodynamics associated with a shear-free congruence of null geodesics.  When stationary and axisymmetric, our solutions reduce to those of Menon and Dermer, the only previously known solutions in Kerr.  In Kerr, all of our solutions have null electromagnetic fields ($\vec{E} \cdot \vec{B} = 0$ and $E^2=B^2$).  However, in Schwarzschild or flat spacetime there is freedom to add a magnetic monopole field, making the solutions magnetically dominated ($B^2>E^2$).  This freedom may be used to reproduce the various flat-spacetime and Schwarzschild-spacetime (split) monopole solutions available in the literature (due to Michel and later authors), and to obtain a large class of time-dependent, non-axisymmetric generalizations.  These generalizations may be used to model the magnetosphere of a conducting star that rotates with arbitrary prescribed time-dependent rotation axis and speed.  We thus significantly enlarge the class of known exact solutions, while organizing and unifying previously discovered solutions in terms of their null structure.

\end{abstract}

\maketitle

\section{Introduction}

Four decades ago, seminal work on pulsars \cite{goldreich-julian1969} and active galactic nuclei \cite{blandford-znajek1977} established the equations of force-free electrodynamics on compact object spacetimes as a basic tool for understanding some of the most dramatic of observed astrophysical phenomena. Since that time, only a handful of exact solutions have been discovered \cite{michel1973b,blandford1976,menon-dermer2007,lyutikov2011}, with only a single solution known in the Kerr black hole spacetime \cite{menon-dermer2007}.  While much progress has been made with numerical simulation of the force-free equations (e.g., \cite{contopoulos-kazanas-fendt1999,komissarov2001,uzdensky2005,spitkovsky2006,mckinney2006,timokhin2006,palenzuela-lehner-liebling2010,alic-etal2012,kalapotharakos-contopoulos-kazanas2012}), many basic questions remain, for which analytic solutions could be helpful.  Exact solutions can provide clean examples of basic physical processes and expose subtleties not easily recognized in simulations of finite resolution.  They may also be used as backgrounds for perturbation theory and as benchmarks for numerical studies.  The equations of force-free electrodynamics are non-linear, but, compared to, say, those of general relativity, not particularly complex.  Encouraged by the fifty-year gap between the invention of general relativity and the discovery of the Kerr metric, we take the optimistic view that it would be too early to give up the search for exact force-free solutions.

The present contribution to this search takes its cue from two known solutions: the only known solution in Kerr, that of Menon and Dermer \cite{menon-dermer2007}, and, what is perhaps the most important solution in flat spacetime, Michel's rotating monopole magnetosphere \cite{michel1973b}.  After finding their solution in a 3+1 approach, Menon and Dermer noted \cite{menon-dermer2009} that its charge-current four-vector was proportional to a principal null\footnote{A null current is composed, in any frame, of charge and three-current densities of equal magnitude. If charged particles of both signs are present, a null current need not involve particles moving with the velocity of light.} congruence of Kerr.  In fact, the Michel monopole solution shares the same property in flat spacetime (the four-current is radial and null), providing a link between these solutions. Michel expressed astonishment that his simple, exact solution [Eq.~(26) therein]  emerged from the general, nonlinear equation for the Euler potential function [Eq.~(11) therein].  Similarly, Menon and Dermer considered a 3+1 version of the non-linear stream equation first derived thirty years prior by Blandford and Znajek \cite{blandford-znajek1977}, and found a simple solution.  In the language in which they are derived, there is no apparent connection between these solutions, no clear reason \textit{why} they exist, and no obvious route to finding generalizations.  In this paper we will assume from the outset that the current is null.  Using a computational method that is well-adapted to the null congruence, we will easily recover these solutions as part of a wide class
that includes time-dependent and non-axisymmetric solutions.

Our basic tool will be the spin-coefficient formalism of Newman and Penrose \cite{newman-penrose1962}, where one works with the scalar projections of the field strength and four-current onto a null tetrad.  Within this framework, a dramatic simplification occurs when the current is assumed to be along one of the (real) null vectors in the tetrad: The force-free equations are reduced from a nonlinear, coupled system of equations for six real functions to a linear system of equations for three real functions that is only mildly coupled.  This simplified system immediately gives rise to some general results relating the existence of force-free solutions to the expansion, shear, and twist of the null congruence.  We integrate the equations for the specific case of the principal null congruence(s) of Kerr, finding a large class of solutions specified in terms of a free real function of three variables. In the stationary and axisymmetric case we reproduce the solution of Menon and Dermer, and establish that it is the unique stationary and axisymmetric solution with current along a principal null direction.  

The electromagnetic field strength of the Menon-Dermer solution is null in the sense that $B^2=E^2$, and our generalizations in Kerr share this property.  It is generally assumed that a physical force-free solution should be magnetically dominated, $B^2 >E^2$.  A main reason for this is the expectation that free charges would quickly screen a strong, electrically dominated field.  Additionally, the equations are no longer hyperbolic when $B^2<E^2$, i.e., there is no well-posed initial-value problem for electrically dominated fields.  The physical relevance of the boundary null case, $E^2=B^2$, is less clear.  As Menon and Dermer's solutions demonstrate, there exist force free field configurations that are null everywhere.  While the stability of these solutions has not yet been investigated, at present we see no physical reason forbidding the realization of a null force-free solution.

Even if they ultimately turn out not to be viable as global models for the environment of a compact object, there is good reason for studying null solutions.  As is well known, the equations of force-free electrodynamics do not preserve the condition of magnetic domination under time evolution.  Generically $B^2-E^2$ becomes smaller in some regions until it crosses zero.  At this point, numerical simulations employ various prescriptions---all \textit{ad hoc}---to reset the field back to acceptable values, often simply returning it to the null value $E^2=B^2$ \cite{spitkovsky2006}.  For this reason null fields play a central role even in numerical evolutions that enforce magnetic domination.  Understanding null exact solutions could shed light on these procedures and help determine their validity.

The Michel monopole solution also has a null current, but is magnetically dominated rather than null.  What is the connection to the null-current solutions studied here?  When the black hole spin parameter is zero (i.e., for Schwarzschild or flat spacetime), our analysis reveals a freedom to add in a (vacuum) magnetic monopole field while preserving the force-free condition.  This monopole field may be split 
into two domains with oppositely directed magnetic field, as a crude model of a dipole, with a current sheet on the interface. In the stationary, axisymmetric case in flat spacetime, adding the monopole reproduces the classic magnetosphere of Michel \cite{michel1973b}, as generalized to arbitrary polar angle dependence by Blandford and Znajek \cite{blandford-znajek1977}.  In the axisymmetric case, doing so reproduces the recent time-dependent solution due to Lyutikov \cite{lyutikov2011}.  More generally, we produce in this way a large class of magnetic, time-dependent and/or non-axisymmetric solutions in Schwarzschild and flat spacetime.  In a forthcoming publication we will show how these solutions can be used to model the magnetosphere of a split-monopole-magnetized conducting star whose rotation axis and speed may change with time.

The time-dependent solutions we have found include (non-linear) waves propagating on the black hole background. Remarkably, these waves preserve the flat-space property of being either purely ingoing or purely outgoing, i.e., \textit{the black hole does not scatter the waves}.  That such behavior is possible is very surprising, given that wave scattering by curvature is generic.  The lack of scattering can be understood in terms of a property of vacuum electrodynamics discovered by Robinson in 1961 \cite{robinson1961}, namely that associated to every (analytic \cite{sommers1976}) shear-free congruence of null geodesics there is a corresponding \textit{local}, null solution to the vacuum Maxwell equations that propagates along the congruence without scattering.  We show explicitly that for the principal congruences of Kerr such vacuum solutions cannot be extended globally without singularities,  a result that also follows from the general analysis of Teukolsky and Press \cite{teukolsky-press1974}. However, in the force-free case, where plasma charge-current supports the electromagnetic fields, global solutions can (and do) exist, giving rise to \textit{bona fide} non-scattering waves.

In section \ref{sec:ffnp} we introduce the equations of force-free electrodynamics, express them in the Newman-Penrose (NP) formalism, and obtain a few general results for null currents.  In section \ref{sec:kerr} we specialize to the principal null congruences of Kerr and display the equations in a compact form.  In section \ref{sec:ingoing} we solve the equations for current along the ingoing congruence, and in section \ref{sec:outgoing} we explain how the corresponding outgoing solutions may be obtained.   In section \ref{sec:magnetospheres} we briefly discuss the role of these solutions in modeling pulsar magnetospheres. An appendix relates the Newman-Penrose description to ordinary electric and magnetic field components in flat spacetime.  We adopt the signature $(+,-,-,-)$, and use Gaussian units with $G=c=1$.

\section{Force-free Electrodynamics and the Newman-Penrose Framework}\label{sec:ffnp}

Classical charged matter is described by Maxwell's equations,
\begin{align}
\nabla_\nu F^{\mu \nu} & = 4 \pi J^\mu \label{maxmat1} \\
\nabla_{[\mu} F_{\nu \rho]} & = 0, \label{maxmat2} 
\end{align}
together with conservation of total stress-energy, 
\begin{equation}\label{maxmat3}
\nabla^\nu T^{EM}_{\mu \nu} = -F_{\mu\nu}J^\nu = -\nabla^\nu  T_{\mu \nu}. 
\end{equation}
Here $F_{\mu \nu}$ is the field strength tensor, $J^\mu$ is the four-current, $T_{\mu \nu}$ is the matter stress-energy, and $T^{EM}_{\mu \nu}$ is the electromagnetic field stress-energy,
\begin{equation}
4 \pi T^{EM}_{\mu \nu} = F_{\mu \alpha} F_{\nu}{}^{\alpha} - \tfrac{1}{4} g_{\mu \nu} F_{\alpha \beta} F^{\alpha \beta}. 
\end{equation}
For each particular type of matter, these equations would be supplemented with 
information about the matter dynamics to provide a deterministic set of equations.  Equation \eqref{maxmat3} describes the exchange of energy and momentum between the matter and the electromagnetic field via the Lorentz force density $F_{\mu\nu}J^\nu$.  

Force-free electrodynamics governs the dynamics of the field and current when the energy and momentum stored in the matter is far smaller than that in the field.  (For example, consider a low-density plasma in a strong magnetic field, such that the rest mass density is small compared with the magnetic energy density.)  In this case it should be a good approximation to neglect the transfer of energy and momentum entirely, leading to the \textit{force-free condition},
\begin{equation}
F_{\mu \nu} J^\nu = 0. \label{ff}
\end{equation}
Using the Maxwell equation (\ref{maxmat1}) to eliminate the current, 
the force-free condition can be expressed purely in terms of the field, without reference to the matter, as $F_{\mu \nu} \nabla_\rho F^{\nu \rho}=0$.
In this way the matter is removed entirely from the description. The resulting system is described by the equations of force-free electrodynamics,
\begin{align}
\nabla_\nu F^{\mu \nu} & = 4 \pi J^\mu \label{ff1} \\
\nabla_{[\mu} F_{\nu \rho]} & = 0 \label{ff2} \\
F_{\mu \nu} J^\nu & = 0. \label{ff3}
\end{align}
We may regard the field strength $F_{\mu \nu}$ as the only fundamental variable, 
with equation \eqref{ff1} having the status of \textit{defining} of the charge-current. 
Given a solution of the force-free equations, the current is therefore determined,
but there is otherwise no information about what the matter is doing.

From equations \eqref{ff1}-\eqref{ff3} it is far from clear whether force-free electrodynamics has the basic mathematical properties expected of a physical theory, such as a well-posed initial value problem.  To analyze the status of equations \eqref{ff1}-\eqref{ff3} as an evolution problem, one may make a 3+1 split.  Working in flat spacetime, one can show that all solutions to the equations satisfy $\vec{E} \cdot \vec{B}=0$ and $\vec{\jmath} = B^{-2}[(\vec{\nabla}\cdot\vec{E})\vec{E} \times \vec{B} + (\vec B\cdot\vec\nabla\times\vec B -\vec E\cdot\vec\nabla\times\vec E)\vec B]$.  One can then ask whether pairing these equations with Maxwell's equations represents a well-posed initial value problem (i.e., gives rise to a hyperbolic system of equations).  It turns out that that hyperbolicity hinges on the sign of the invariant $F^2=F_{\mu \nu}F^{\mu \nu} \propto B^2-E^2$ \cite{komissarov2002}.  For magnetic domination, $F^2>0$, a symmetric hyperbolic formulation is possible \cite{pfeiffer-macfadyen}, guaranteeing well-posed evolution.  For electric domination, $F^2<0$, the equations are not strongly hyperbolic, and evolution is ill-posed.  The marginal case $F^2=0$ has not been analyzed.  The above comments should generalize readily to curved spacetime.

\subsection{Newman-Penrose Framework}

Our analysis in this paper will make use of the NP framework \cite{newman-penrose1962}.  This framework involves a tetrad of null vectors \{$\ell^\mu,n^\mu,m^\mu,\bar{m}^\mu\}$, where 
$\ell^\mu$ and $n^\mu$ are real, while $m^\mu$ is complex (with the overbar denoting complex conjugation).  The vectors must satisfy $\ell_\mu n^\mu=1$ and $m_\mu \bar{m}^\mu=-1$, with all other inner products vanishing.  The metric may be reconstructed from the tetrad by
\begin{equation}
g^{\mu \nu} = \ell^\mu n^\nu + n^\mu \ell^\nu - m^\mu \bar{m}^\nu - \bar{m}^\mu m^\nu.
\end{equation}
This framework is particularly useful when one or two real null directions are singled out 
by the problem at hand. Note that even then there remains the freedom to rescale the vectors
as $\ell^\mu\rightarrow \lambda\ell^\mu$, $n^\mu\rightarrow \lambda^{-1}n^\mu$, 
$m^\mu\rightarrow e^{i\theta}m^\mu$, with $\lambda$ and $\theta$ real.
In lieu of the Christoffel symbols one works with covariant derivatives of the tetrad vectors projected on to the tetrad vectors, or ``spin coefficients''.  There are twelve such complex spin coefficients, each given its own individual Greek letter.

To express the electromagnetic field in the NP framework one defines complex scalars $\phi_0,\phi_1,\phi_2$ as follows,
\begin{align}
\phi_0 & = F_{\mu \nu} \ell^\mu m^\nu \label{phi0} \\
\phi_1 & = \tfrac{1}{2} F_{\mu \nu} \left( \ell^\mu n^\nu + \bar{m}^\mu m^\nu \right) \label{phi1} \\
\phi_2 & = F_{\mu \nu} \bar{m}^\mu n^\nu. \label{phi2}
\end{align}
One thus replaces the six real components of the antisymmetric tensor $F_{\mu \nu}$ with three complex scalar fields.  The field strength is reconstructed from the scalars via
\begin{equation}\label{Fphi}
F_{\mu \nu}=2[\phi_1(n_{[\mu}\ell _{\nu]}+m_{[\mu}\bar{m}_{\nu]}) + \phi_2\ell_{[\mu}m_{\nu]} + \phi_0\bar{m}_{[\mu}n_{\nu]}] + \textrm{c.c.},
\end{equation}
where c.c. denotes the complex conjugate of the preceding terms. Maxwell's equations then become \cite{newman-penrose1962,teukolsky1973}
\begin{align}
\begin{split}
(\ell \cdot\! \nabla -2 \rho) \phi_1-(\bar{m} \cdot\! \nabla+\pi-2 \alpha) \phi_0 & = - \kappa \phi_2 +  2 \pi J_{\ell}\\
-(n \cdot\! \nabla+2 \mu)\phi_1 + (m \cdot\! \nabla-\tau+2\beta)\phi_2 & = -\nu \phi_0 + 2 \pi J_n\\
(m \cdot\! \nabla- 2 \tau) \phi_1-(n \cdot\! \nabla+ \mu - 2\gamma)\phi_0 & = -\sigma \phi_2 + 2 \pi J_m\\
-(\bar{m} \cdot\! \nabla+2\pi)\phi_1 + (\ell \cdot\! \nabla - \rho +2 \epsilon)\phi_2 & = -\lambda \phi_0+ 2 \pi J_{\bar{m}}.
\end{split}\label{maxnp}
\end{align}
Here $\{\alpha,\beta,\gamma,\epsilon,\kappa,\lambda,\mu,\nu,\pi,\rho,\sigma,\tau\}$ are the twelve spin coefficients, defined in \cite{newman-penrose1962}, while the $J$ with subscript letters indicate projections, $J_{\ell}=J^\mu \ell_\mu$, $J_{n}=J^\mu n_\mu$ and $J_m=J^\mu m_\mu$.  (Note that the $\pi$ multiplying the current is the number, rather than the spin coefficient.)  The $\cdot$ notation indicates contraction, e.g., $\ell \cdot\! \nabla = \ell^\mu \nabla_\mu$. 

The quadratic invariants are given in NP language by
\begin{align}
\tfrac{1}{2} F_{\mu \nu} F^{\mu \nu} & = 4 \textrm{Re}[\phi_0 \phi_2 - \phi_1^2]= B^2 - E^2 \label{F2} \\
 \tfrac{1}{4} \epsilon_{\mu \nu \lambda \rho}F^{\mu \nu}F^{\lambda \rho} & = 2 \textrm{Im}[\phi_0 \phi_2 - \phi_1^2]=\vec{E} \cdot \vec{B}.\label{FwF}
\end{align}
As discussed above, the first invariant, equation \eqref{F2}, controls the hyperbolicity of the force-free equations, while the second invariant, equation \eqref{FwF}, will vanish for any force-free field.  Finally, the force-free condition \eqref{ff3} is given by
\begin{align}
\textrm{Re}[\phi_1 J_n - \phi_2 J_m] & = 0 \label{ffnp1} \\
\textrm{Re}[\phi_1 J_\ell - \bar{\phi}_0 J_m] & = 0 \label{ffnp2} \\
2 i \textrm{Im}[\phi_1] J_{\bar{m}} + \bar{\phi}_0 J_n - \phi_2 J_\ell & = 0. \label{ffnp3}
\end{align}

\subsection{Null current}
\label{sec:nullcurrent}

The NP version of the force-free equations becomes very simple when the current is taken along one of the null directions, $\ell^\mu$ or $n^\mu$.  For example, if $J^\mu \propto n^\mu$, then the force-free condition becomes $F_{\mu \nu} n^\mu=0$ and it follows immediately from equations \eqref{phi0}-\eqref{phi2} that $\phi_2$ vanishes and $\phi_1$ is purely imaginary,
\begin{equation}
\phi_2=0 \textrm{ and } \textrm{Re}[\phi_1]=0 \qquad \textrm{ (FF condition when } J^\mu \propto n^\mu \textrm{)}.\label{ffnpnull}
\end{equation}
(A similar simplification occurs for $J^\mu \propto \ell^\mu$: $\phi_0$ must vanish and $\phi_1$ must be purely imaginary.)  This simplification may also be seen from equations \eqref{ffnp1}-\eqref{ffnp3} with $J_m=J_n=0$.  When the conditions (\ref{ffnpnull}) hold, $B^2-E^2$ (\ref{F2}) is positive unless $\phi_1=0$, in which case it vanishes. So force-free fields with null current along $n^\mu$ are null if $\phi_1=0$ and magnetic if
$\phi_1\ne0$, but are never electric. The field strength (\ref{Fphi}) takes the form 
\beq
F_{\mu\nu}= 2\phi_1\, m_{[\mu}\bar{m}_{\nu]} + 2 \phi_0\, \bar{m}_{[\mu}n_{\nu]} + c.c.
\eeq
Some intuition for this form may be gained by considering the orthonormal tetrad $(T,X,Y,Z)$ defined by $\sqrt{2}T=l+n$, $\sqrt{2}Z=l-n$, 
$\sqrt{2}X=m+\bar{m}$, $-i\sqrt{2}Y= m-\bar{m}$.  We then have a null current in the $Z$-direction, and a field
\beq
F_{\mu\nu}= 4\textrm{Im}[\phi_1]\, X_{[\mu}Y_{\nu]} + 
2(\textrm{Re}[\phi_0] X + \textrm{Im}[\phi_0] Y)_{[\mu}(T-Z)_{\nu]}.
\eeq
The first term is a magnetic field in the $Z$-direction, while the second term is a null field consisting
of perpendicular electric and magnetic fields of equal magnitude, transverse to the $Z$-direction.

With $\phi_2=0$ and $J_n=J_m=0$,  and Maxwell's equations (\ref{maxmat1},\ref{maxmat2}) become 
\begin{align}
(\ell \cdot\! \nabla - 2 \rho) \phi_1-(\bar{m} \cdot\! \nabla+\pi-2 \alpha) \phi_0 & = 2 \pi J_{\ell} \label{potato1} \\
-(n \cdot\! \nabla+2 \mu)\phi_1 & = -\nu \phi_0 \label{potato2} \\
(m \cdot\! \nabla- 2 \tau) \phi_1-(n \cdot\! \nabla+ \mu - 2\gamma)\phi_0 & = 0 \label{potato3} \\
-(\bar{m} \cdot\! \nabla+2\pi)\phi_1 & = -\lambda \phi_0 . \label{potato4}
\end{align}
If the solution is null ($\phi_1=0$), then equations \eqref{potato2} and \eqref{potato4} require 
$\nu =\lambda =0$, which implies that the congruence is geodesic and shear-free \cite{newman-penrose1962}. Now suppose that the solution is magnetic ($\phi_1\ne0$), and that the $n^\mu$ congruence is geodesic, so that $\nu=0$. Then, since $\phi_1$ is imaginary (\ref{ffnpnull}), the real part of equation \eqref{potato2} requires that $\textrm{Im}[\mu]=0$, which implies that the congruence must be twist-free \cite{newman-penrose1962}.  We have thus established the following properties of solutions to force-free electrodynamics with null current:
\begin{itemize}
\item[i)] There are no electric solutions.
\item[ii)] Null solutions can exist only if the the congruence is geodesic and shear free.
\item[iii)] Magnetic solutions with null geodesic current can exist only if the congruence is twist-free, i.e., hypersurface-orthogonal.
\end{itemize}
Property (ii) holds even when the current vanishes, and in that case it is the `only if' part of Robinson's theorem \cite{robinson1961,sommers1976}, which asserts that null solutions to the vacuum Maxwell equations exist locally if and only if the associated null congruence is geodesic and shear-free. (Analyticity is required for the `if' direction.)  A null-current Robinson theorem has been given previously by Zund \cite{zund1974}.\footnote{The study of null-current sources appears to have begun 100 years ago with the work of Bateman \cite{bateman1913}, who described his source as a ``double-barreled gun G which moves through the aether in an arbitrary manner and keeps up a perpetual fire of secondary singularities (bullets)...''.  Bateman later dubbed such sources ``light darts'' \cite{bateman1923}.}

\section{Fields in the Kerr spacetime}\label{sec:kerr}

In flat spacetime, in spherical coordinates, the null vector field $\ell^\mu=(1,1,0,0)$ represents light rays emitted from the origin and moving radially outward (approaching ``future null infinity''), while the null vector field $n^\mu=(1,-1,0,0)$ represents light rays emitted far away, in the distant past (``past null infinity'') and converging to the origin.  A natural generalization of these congruences to the Kerr spacetime are the so-called principal null congruences, picked out by their special relationship to the algebraic properties of the Weyl curvature tensor according to the Petrov classification (e.g., \cite{wald-book}).  Here the role of the origin is replaced by the event horizon(s); the outgoing congruence $\ell^\mu$ represents light rays emerging from the white hole (out of the past horizon) and traveling to future null infinity, while the ingoing congruence $n^\mu$ represents light rays emitted from past null 
infinity and traveling into the black hole (through the future horizon).  These congruences are geodesic and shear-free, so that the NP spin coefficients 
$\{\kappa,\nu,\sigma,\lambda\}$ are zero.  With these congruences, Maxwell's equations (\ref{maxnp}) take the simpler form
\begin{align}
\begin{split}
(\ell \cdot\! \nabla - 2 \rho) \phi_1-(\bar{m} \cdot\! \nabla+\pi-2 \alpha) \phi_0 & = 2 \pi J_{\ell}\\
-(n \cdot\! \nabla+2 \mu)\phi_1 + (m \cdot\! \nabla-\tau+2\beta)\phi_2 & =  2 \pi J_n\\
(m \cdot\! \nabla- 2 \tau) \phi_1-(n \cdot\! \nabla+ \mu - 2\gamma)\phi_0 & = 2 \pi J_m\\
-(\bar{m} \cdot\! \nabla+2\pi)\phi_1 + (\ell \cdot\! \nabla - \rho +2 \epsilon)\phi_2 & = 2 \pi J_{\bar{m}}.
\end{split}\label{maxnpkerr}
\end{align}

We will use the standard Kinnersley tetrad \cite{kinnersley1969}, which has the property that $\epsilon$ vanishes, so that in particular the outgoing congruence is affinely parameterized (while the ingoing congruence is not).  Explicitly, the Kinnersley tetrad is given in Boyer-Lindquist coordinates $(t,r,\theta, \phi)$ by
\begin{align}
\ell^\mu & =\frac{1}{\Delta} \left(r^2+a^2, \Delta, 0,a \right) \label{ellbl} \\
n^\mu & = \frac{1}{2 \Sigma} \left( r^2 + a^2, -\Delta, 0, a \right) \label{nbl} \\
m^\mu & = \frac{- \bar{\rho}}{\sqrt{2}} \left( i a \sin \theta, 0, 1, \frac{i}{\sin \theta} \right), \label{mbl}
\end{align}
where
\begin{align}
\Delta & = r^2 + a^2 - 2 M r \label{deltadef} \\
\Sigma & = r^2 + a^2 \cos^2 \theta \label{sigmadef} \\
\rho & = \frac{-1}{r- i a \cos \theta} \label{rhodef}.
\end{align}
These quantities have the following interpretations:  $\Delta=0$ gives the position of the horizon(s), the metric volume element is given by $\Sigma \sin \theta d^4 x$, and $\rho=\bar{m}^a m^b \nabla_a \ell_b$ is the NP spin coefficient that measures the expansion and twist of the congruence defined by $\ell^\mu$.  (By the symmetries of Kerr, this also gives the expansion and twist of the ingoing congruence.) Note also the useful relationship that $\Sigma^{-1} = \rho \bar{\rho}$.

Expressions for all of the Kinnersley tetrad spin coefficients in Boyer-Lindquist coordinates are given in \cite{teukolsky1973}.  After substituting these for the spin coefficients in equations \eqref{maxnpkerr}, we notice that the equations can be simplified by introducing certain rescaling factors.  In particular, we may write Maxwell's equations in the Kerr metric as
\begin{align}
\rho^2 \ell \cdot\! \nabla (\rho^{-2} \phi_1) - \frac{1}{\alpha_0}\bar{m} \cdot\! \nabla ( \alpha_0 \phi_0 ) & = 2 \pi J_\ell \label{maxnpkerrsimp1} \\
-\rho^2 n \cdot\! \nabla (\rho^{-2} \phi_1) + \frac{1}{\alpha_2}m \cdot\! \nabla ( \alpha_2 \phi_2 ) & = 2 \pi J_n \label{maxnpkerrsimp2} \\
\rho^2 m \cdot\! \nabla (\rho^{-2} \phi_1) - \frac{1}{\alpha_0}n \cdot\! \nabla ( \alpha_0 \phi_0 ) & = 2 \pi J_m \label{maxnpkerrsimp3} \\
-\rho^2 \bar{m} \cdot\! \nabla (\rho^{-2} \phi_1) + \frac{1}{\alpha_2}\ell \cdot\! \nabla ( \alpha_2 \phi_2 ) & = 2 \pi J_{\bar{m}}, \label{maxnpkerrsimp4}
\end{align}
where
\begin{align}
\alpha_0 & = \Delta \rho \sin \theta \label{alpha0}\\
\alpha_2 & = \rho^{-1} \sin \theta. \label{alpha2}
\end{align}
Note that this rescaling of the NP scalars is not associated with a rescaling of the tetrad, since $\phi_1$ would be invariant under that operation, 
and the scalings of $\phi_0$ and $\phi_2$ would be reciprocal. We do not have any clear insight into why a rescaling that absorbs all non-derivative terms in (\ref{maxnpkerr}) is possible. 

Equations \eqref{maxnpkerrsimp1}-\eqref{maxnpkerrsimp4} are the basic starting point for our analysis of Maxwell's equations in Kerr.   While we will not use these equations in their full generality, we present them in case they are useful to future researchers.  No approximations have been used in writing equations \eqref{maxnpkerrsimp1}-\eqref{maxnpkerrsimp4}, and the force-free condition has not been applied; equations \eqref{maxnpkerrsimp1}-\eqref{maxnpkerrsimp4} are simply a rewriting of Maxwell's equations.

\subsection{Ingoing Coordinates}

While their derivation made use of Boyer-Lindquist coordinates, equations \eqref{maxnpkerrsimp1}-\eqref{maxnpkerrsimp4} are given in a covariant form, so that one can straightforwardly use other coordinate systems.  We will find the ingoing Kerr coordinate system, which is regular on the future horizon, particularly useful (see \cite{poisson-book} for a nice treatment).  
The ingoing coordinates $(v,r,\theta,\psi)$ are defined by 
\begin{align}
v & = t+ r^* \\
\psi & = \phi + r^\sharp,
\end{align}
where $r^*$ and $r^\sharp$ satisfy
\begin{align}
\frac{d r^*}{dr} & = \frac{r^2+a^2}{\Delta} \\
\frac{d r^\sharp}{dr} & = \frac{a}{\Delta},
\end{align}
and are given explicitly by
\begin{align}
r^* & = r + \frac{M r_+}{\sqrt{M^2-a^2}} \log \left| \frac{r}{r_+} - 1 \right| - \frac{M r_-}{\sqrt{M^2-a^2}} \log \left| \frac{r}{r_-}-1 \right| \\
r^\sharp & = \frac{a}{2 \sqrt{M^2-a^2}} \log \left| \frac{r-r_+}{r-r_-} \right|.
\end{align}
Here $r_{+}(r_{-})$ is the outer(inner) horizon coordinate radius,
\begin{equation}
r_{\pm} = r \pm \sqrt{M^2-a^2}.
\end{equation}
In these coordinates the tetrad vector components become
\begin{align}
\ell^\mu & = \left( 2\frac{r^2+a^2}{\Delta}, 1, 0, 2\frac{a}{\Delta} \right) \label{ellin} \\
n^\mu & =  \left( 0, -\frac{\Delta}{2 \Sigma}, 0, 0 \right) \label{nin} \\
m^\mu & = \frac{- \bar{\rho}}{\sqrt{2}} \left( i a \sin \theta, 0, 1, \frac{i}{\sin \theta} \right). \label{min}
\end{align}
A key simplification is that the ingoing vector now has a single non-zero component, $n \propto \partial_r$.  The ingoing coordinates are regular on the future (black hole) horizon, but singular on the past (white hole) horizon.

\subsection{Outgoing coordinates}

The outgoing Kerr coordinates $(u,r,\theta,\chi)$ are defined by (e.g., \cite{poisson-book})
\begin{align}
u & = t - r^* \\
\chi & = \phi - r^\sharp,
\end{align}
and the Kinnersley tetrad components are
\begin{align}
\ell^\mu & = \left( 0, 1, 0, 0 \right)  \label{ellout} \\
n^\mu & = \frac{1}{2 \Sigma}\Bigl(2(r^2+a^2), -\Delta, 0, 2a \Bigr) \label{nout} \\
m^\mu & = \frac{- \bar{\rho}}{\sqrt{2}} \left( i a \sin \theta, 0, 1, \frac{i}{\sin \theta} \right). \label{mout}
\end{align}
These coordinates are regular on the past (white hole) horizon, but singular on the future horizon.

\subsection{Conditions for regularity at the horizons}

Equations \eqref{ellin}-\eqref{min} reveal that the ingoing Kerr coordinate components of the Kinnersley tetrad are singular at $\Delta=0$.  Since the coordinates are well behaved, the Kinnersley tetrad itself is not regular on the future horizon.  However, we may produce a regular tetrad by performing the rescaling $\ell^\mu \rightarrow \Delta \ell^\mu, n^\mu \rightarrow \Delta^{-1} n^\mu$, which preserves the condition $\ell^\mu n_\mu=1$.  From equations \eqref{phi0}-\eqref{phi2}, the NP scalars transform as $\phi_0 \rightarrow \Delta \phi_0$ and $\phi_2 \rightarrow \Delta^{-1} \phi_2$, with $\phi_1$ unchanged, and it is these new scalars that must be regular at $\Delta=0$.  Thus the condition for regularity of the solution on the future horizon is
\begin{equation}
\textrm{Future horizon: } \{ \Delta \phi_0, \phi_1, \Delta^{-1} \phi_2\} \textrm{ must be regular}. \label{futurereg}
\end{equation}
For the past horizon we may use the outgoing Kerr coordinates.  Equations \eqref{ellout}-\eqref{mout} reveal that the Kinnersley tetrad is regular on the past horizon, and so the regularity conditions are just
\begin{equation}
\textrm{Past horizon: } \{\phi_0, \phi_1, \phi_2\} \textrm{ must be regular.} \label{pastreg}
\end{equation}

\section{Ingoing Solutions}\label{sec:ingoing}

We begin with the ingoing case, i.e., solutions whose current runs along the ingoing null congruence,
\begin{equation}\label{Jn}
J^\mu = \mathcal{J} n^\mu.
\end{equation}
In particular we have $J_n = J_m = 0$ and $J_\ell = \mathcal{J}$.
According to equation \eqref{ffnpnull}, the force-free condition becomes simply
\begin{equation}
\phi_2=0 \textrm{ and } \textrm{Re}[\phi_1]=0.\label{ffnpnull2}
\end{equation}
Working in ingoing coordinates $(v,r,\theta,\psi)$, we will find that  $\phi_0 \propto r^{-1}$ and $\phi_1\propto r^{-2}$ for large $r$.  The stress tensor then approaches the form $2\pi T^{EM}_{\mu \nu} = |\phi_0|^2 n_\mu n_\nu$ for large $r$, representing an ingoing energy flux.  In particular, the flux of Killing energy through a sphere at $r=\infty$ is given by\footnote{This limit is at fixed $v$, approaching past null infinity.  In the stationary case flux integrals are often computed at spatial infinity, $r \rightarrow \infty$ fixing $t$.  This is equivalent to our \eqref{Efluxin} for stationary solutions.}
\beq\label{Efluxin}
\lim_{r \rightarrow \infty} \int (T^{EM})^r{}_v r^2 d \Omega =-\frac{1}{8\pi} \lim_{r \rightarrow \infty} \int |\phi_0|^2 r^2 d\Omega < 0,
\eeq
where $d\Omega$ is the two-sphere area element. Although the energy flux is always inward,  the three-current associated with (\ref{Jn}) is outgoing where $\mathcal{J}<0$.

With the null current (\ref{Jn}), Maxwell's equations (\ref{maxnpkerrsimp1}-\ref{maxnpkerrsimp4}) are just
\begin{align}
\rho^2 \ell \cdot\! \nabla (\rho^{-2} \phi_1) - \frac{1}{\alpha_0}\bar{m} \cdot\! \nabla ( \alpha_0 \phi_0 ) & = 2 \pi \mathcal{J} \label{maxin1} \\
n \cdot\! \nabla (\rho^{-2} \phi_1)  & = 0  \label{maxin2} \\
\rho^2 m \cdot\! \nabla (\rho^{-2} \phi_1) - \frac{1}{\alpha_0}n \cdot\! \nabla ( \alpha_0 \phi_0 ) & = 0 \label{maxin3} \\
\bar{m} \cdot\! \nabla (\rho^{-2} \phi_1) & = 0. \label{maxin4}
\end{align}
Equation (\ref{maxin2}) implies that $\rho^{-2}\phi_1$ is independent of $r$. Since $\phi_1$ is 
pure imaginary, and $\rho^{-2}=(r-ia\cos\theta)^2$ is complex and $r$-dependent, this is impossible
unless $a=0$. This is an example of the general result established earlier in Sec. \ref{sec:nullcurrent}:
$\phi_1$ must vanish if the null current congruence is twisting, as it is in Kerr.

\subsection{Solution for $\phi_1$}

To determine the general solution for $\phi_1$ we may thus restrict to Schwarzschild spacetime.  Here the ingoing Kerr coordinates $(v,r,\theta,\psi)$ become ingoing Eddington-Finklestein coordinates $(v,r,\theta,\phi)$ (note $\psi=\phi$).  Using $\rho=-1/r$ in Schwarzschild, equations \eqref{ffnpnull2}, \eqref{maxin2} and \eqref{maxin4} imply that $\phi_1$ must be of the form
\begin{equation}
\phi_1 = \frac{iQ(v)}{r^2},
\end{equation}
for some real function $Q$.   
This is a time-dependent magnetic monopole, but Faraday's law does not allow a regular field to have a time-dependent monopole charge.  For a general field with $\phi_1$ given as above, the flux of the field strength 2-form $F$ through a constant $v,r$ sphere is proportional to $Q(v)$.  But the difference in flux at two different values of $v$ is equal to the integral $dF$ over the 3-tube connecting the two spheres. This vanishes since $dF=0$ (which is Faraday's law when pulled back to a timelike 3-surface).  The general solution for $\phi_1$ is therefore the magnetic monopole field
\begin{equation}
\phi_1 = \begin{cases} \displaystyle{\frac{-iq}{2r^2}}, & \textrm{if } a=0  \\ 0, & \textrm{otherwise}, \end{cases}\label{phi1soln}
\end{equation}
where the real constant $q$ is the monopole charge.  This is a source-free solution to all of Maxwell's equations by itself, so it drops out of Eqs. (\ref{maxin1}) and (\ref{maxin3}), leaving an uncoupled equation for $\phi_0$.  

\subsection{Solution for $\phi_0$}

With $\phi_2$ vanishing, and $\phi_1$ given by equation \eqref{phi1soln}, 
Maxwell's equations \eqref{maxin1}-\eqref{maxin4} reduce to just 
\begin{align}
- \frac{1}{\alpha_0}\bar{m} \cdot\! \nabla ( \alpha_0 \phi_0 ) & = 2 \pi \mathcal{J} \label{cutiepie1} \\
n \cdot\! \nabla ( \alpha_0 \phi_0 ) & = 0, \label{cutiepie2}
\end{align}
where $\alpha_0=\Delta \rho \sin \theta$ [see equation \eqref{alpha0}].  The latter equation is simply $\partial_r(\alpha_0 \phi_0)=0$ (where $\partial_r$ is at fixed $v,\theta,\psi$), so that $\phi_0$ takes the form
\begin{equation}
\phi_0 = \frac{f(v,\theta,\psi)}{\Delta \rho}, \label{solved1}
\end{equation}
with some complex function $f$.  Notice that $\phi_0$ diverges as the future horizon is reached, $\Delta \rightarrow 0$ at fixed $v,\theta,\psi$.  However, this divergence is due only to the singular tetrad; we satisfy the regularity condition that $\Delta \phi_0$ remains finite (equation \eqref{futurereg}), and the solution is regular at the future horizon.  The condition for past horizon regularity is that $\phi_0$ itself remain finite as the past horizon is reached ($\Delta \rightarrow 0$ at fixed $u,\theta,\psi$), which, if desired, may be accomplished by appropriate choice of $f(v,\theta,\psi)$.  However, in the stationary case $\partial_v f=0$, equivalently $\partial_t f=0$ in Boyer-Lindquist coordinates, all of the ingoing solutions are singular on the past horizon.

Given equation \eqref{solved1}, equation \eqref{cutiepie1} becomes
\begin{equation}
\left[- i a \partial_v + \frac{1}{\sin \theta}\left(\partial_\theta  - \frac{i}{\sin \theta} \partial_\psi \right) \right]\Big( \sin \theta\,  f(v,\theta,\psi) \Big) = 2 \pi \sqrt{2} \Delta \mathcal{J}. \label{socute}
\end{equation}
Notice that solutions will exist only when $\Delta \mathcal{J}=0$ is independent of $r$, i.e., $n \cdot\! \nabla (\Delta \mathcal{J})=0$.  It is easy to check that this integrability condition is precisely that of current conservation for a current of the form (\ref{Jn}). Although we will not need to make use of it here, 
it is worth noting that the angular derivative term in (\ref{socute}) can be identified as
$-\bar\eth f$, where $\eth$ is the spin-weight lowering operator associated with 
the unit sphere \cite{Newman:1966ub}, and $f$ has spin weight +1.  We will now discuss solutions of this equation, in roughly increasing order of complexity.

\subsubsection{Spherically symmetric (Schwarzschild) background}\label{genschw}

In Schwarzschild spacetime, $a=0$, the $v$-derivative term disappears from equation (\ref{socute}), yielding
\begin{equation}
 \frac{1}{\sin \theta}\left(\partial_\theta  - \frac{i}{\sin \theta} \partial_\phi \right)\Big( \sin\theta\,  f(v,\theta,\phi)\Big) = 2 \pi \sqrt{2} \Delta  \mathcal{J}, \label{socuteschw}
\end{equation}
where $\psi=\phi$ has been used.  Since $\mathcal{J}$ is real, the imaginary part of equation \eqref{socuteschw} reads $\partial_\theta (\sin\theta f^I) - \partial_\phi f^R=0$, where $f^R$ and $f^I$ are the real and imaginary parts of $f$.  At each $v$, this can be expressed as $dw=0$ for a 1-form $w=f^R d\theta + \sin\theta f^I d\phi$ on the sphere, which is smooth when the force-free solution \eqref{solved1} is smooth.   Since all smooth, closed 1-forms on the two-sphere are exact, we have $w=dS=\partial_\theta S\, d\theta+ \partial_\phi S\,  d\phi$ for some smooth real function $S(v,\theta,\phi)$.  Thus
$f^R = \partial_\theta S$ and $f^I = \csc\theta\partial_\phi S$, so that
\begin{equation}\label{ethS}
f = \left( \partial_\theta + \frac{i}{\sin \theta} \partial_\phi \right) S(v,\theta,\phi).
\end{equation}
When this is substituted into  \eqref{socuteschw}, the equation becomes
simply Poisson's equation,
\begin{equation}
\nabla^2 S= 2\pi \sqrt{2} \Delta \mathcal{J},
\label{Jschwarz}
\end{equation}
where $\nabla^2 =\csc\theta \partial_\theta(\sin \theta \partial_\theta ) + \csc^2 \theta \partial_\phi \partial_\phi$ is the Laplacian on the unit two-sphere.\footnote{An alternate derivation of equations \eqref{ethS} and \eqref{Jschwarz} makes use of the known properties of the spin-weight raising and lowering operators $\eth$ and $\bar \eth$, and the spin-weighted spherical harmonics \cite{Newman:1966ub}.   From the definition of $\phi_0$, $f$ has has spin-weight +1 on the sphere, and can therefore be written as $f=-\eth S$, where $S$ has spin-weight 0 and is in general complex.  The left-hand side of equation \eqref{socuteschw} is simply $-\bar{\eth} f$, and, since $\bar\eth\eth = \nabla^2$, equation \eqref{Jschwarz} follows.  The reality of $\mathcal{J}$ and $\nabla^2$ then implies that $S$ must in fact be real (up to an irrelevant constant), giving equation \eqref{ethS} as $f=-\eth S$.} In terms of $S$, the NP scalar $\phi_0$ is then
\begin{equation}\label{phischwarz}
\phi_0 = \frac{-r}{\Delta}\left( \partial_\theta  + \frac{i}{\sin \theta} \partial_\phi \right) S(v,\theta,\phi),
\end{equation}
where we have used $\rho=-1/r$. Having satisfied the imaginary part of \eqref{socuteschw}, the real part is satisfied either by choosing $S$ and \textit{defining} the current by \eqref{Jschwarz}, or by choosing the current and solving \eqref{Jschwarz} for $S$.  Note that  since the integral of $\nabla^2$ over the sphere vanishes, there is no solution unless the integral of $\mathcal{J}$ vanishes.  This means there can be no net three-current flowing through any symmetry two-sphere or, equivalently, that the total charge on the sphere must vanish.  However, provided this condition is satisfied, equation \eqref{Jschwarz} determines $S$ up to a constant, giving rise to a unique force-free solution by equation \eqref{phischwarz}.  Equations \eqref{phischwarz} and \eqref{Jschwarz} provide the general smooth force-free solution with ingoing radial null current in Schwarzschild.  

We have thus found a remarkably wide class of non-stationary, non-axisymmetric solutions on a Schwarzschild background, generalizing previously found solutions with more symmetry \cite{michel1973b,blandford-znajek1977,lyutikov2011}. When the monopole charge $q$ is non-zero, these are magnetically dominated fields, and otherwise they are null. A striking property of these solutions is that, although they may have complicated time-dependence, there is no backscattering of waves from the spacetime curvature. In particular, we may regard the freedom to choose $S(v,\theta,\phi)$ as a choice of initial data for the field at past null infinity, $r \rightarrow \infty$.  We are free to take the data to be compactly supported, corresponding to an ``initial wave packet'' of plasma sent towards the black hole.  This packet propagates into the black hole without scattering, with no plasma or electromagnetic radiation reaching future null infinity ($r \rightarrow \infty$ at fixed $v-2r$).

The existence of these solutions is directly connected to Robinson's theorem \cite{robinson1961}. The radial null congruence is geodesic and shear-free, so this theorem guarantees the existence of local vacuum solutions with null field strength. These correspond to the case $\mathcal{J}=0$, so that $S$ must be a harmonic function on the sphere. However, there are no everywhere regular harmonic functions on the sphere except a constant, which generates the zero solution.  Thus global regularity excludes the existence of Robinson solutions in this case. This property also holds in Kerr, as will be shown in section \ref{sec:nonsymm} below. On the other hand, in the presence of the force-free plasma, non-zero current is allowed, permitting the existence of regular solutions. 

\subsubsection{Stationary case in Kerr}

The stationary case in Kerr is quite similar to the general Scwarzschild case, since in both cases the $i a \partial_v$ term drops out of \eqref{socute}. The differences from the Schwarzschild form are that (i) the monopole field $\phi_1$ is not allowed, (ii) the angle $\phi$ is replaced by $\psi$, (iii) $\Delta$ and $\rho$ have the Kerr form, and (iv) $v$ dependence is excluded.  In particular, the general solution is
\begin{equation}\label{phistat}
\phi_0 = \frac{1}{\Delta \rho}\left( \partial_\theta  + \frac{i}{\sin \theta} \partial_\psi \right) S(\theta,\psi),
\end{equation}
with current
\begin{equation}\label{QJstat}
\nabla^2 S = 2\pi \sqrt{2} \Delta \mathcal{J}.
\end{equation}
Here $\nabla^2$ is the Laplacian associated with the ordinary two-sphere metric  with spherical coordinates $(\theta,\psi)$, $\nabla^2 S=\csc\theta \partial_\theta(\sin \theta \partial_\theta S) + \csc^2 \theta \partial_\psi \partial_\psi S$.  Unlike in the Schwarzschild case, this metric is \textit{not} proportional to the induced metric on surfaces of constant $r$ and $v$.  The integrability condition is that $\mathcal{J}$ must vanish when averaged over the $(\theta,\psi$) sphere with the usual spherical-coordinate two-sphere element $d\Omega = \sin \theta d\theta d\psi$.  However, by stationarity and current conservation this condition is equivalent to the covariant requirement that no current flows through the 3-surface generated by \textit{any} topological 2-sphere along the flow of the timelike Killing vector. In particular, no net current flows in from infinity.

Among the stationary solutions in Kerr are the subclass that are also axisymmetric, which are generated by a choice of axisymmetric potential function, $S=S(\theta)$.   In this case our class reduces to that of Menon and Dermer \cite{menon-dermer2007}.  The relationship between their free function $\Lambda(\theta)$ and our $S(\theta)$ is given by $\sqrt{2} a^2 S(\theta) = -4 \Lambda(\theta) \cos \theta / \sin^5\theta$.  Our analysis has revealed the Menon/Dermer solutions to be the unique stationary, axisymmetric ones with current along the ingoing principal null direction.

\subsubsection{Axisymmetric case in Kerr}

In the axisymmetric case, without assuming stationarity, equation \eqref{socute} becomes
\begin{equation}
\left[- i a \partial_v + \frac{1}{\sin \theta}\partial_\theta\right] \Big(\sin\theta\, f(v,\theta) \Big) = 2 \pi \sqrt{2} \Delta \mathcal{J}. \label{socuteaxi}
\end{equation}
Following similar logic to that used below equation \eqref{socuteschw}, we find that, at least locally, the general solution is given by
\begin{equation}
f = \frac{1}{\sin\theta}\left(i a \partial_v+\frac{1}{\sin \theta}\partial_\theta  \right) P(v,\theta)
\end{equation}
for a real function $P(v,\theta)$.  Then the field and current are given by
\begin{align}
\phi_0 & = \frac{1}{\Delta \rho \sin \theta}\left[i a \partial_v P+\frac{1}{\sin \theta}\partial_\theta P \right] \label{axi1} \\
\mathcal{J} & = \frac{1}{2\pi \sqrt{2} \Delta} \left[a^2 \partial_v \partial_v P+ \frac{1}{\sin \theta}\partial_\theta \left( \frac{1}{\sin \theta} \partial_\theta P \right) \right]. \label{axi2}
\end{align}
Any choice of a real function $P(v,\theta)$ gives rise to an axisymmetric solution by equations \eqref{axi1} and \eqref{axi2}.  However, some such solutions will be singular at the poles $\theta=\{0,\pi\}$, which are not covered by our coordinates.

\subsubsection{General non-symmetric case in Kerr}\label{sec:nonsymm}

Finally, we discuss the completely general case. The real and imaginary parts of equation \eqref{socute} are
\begin{align}
a \sin \theta \, \partial_v f^I + \frac{1}{\sin \theta}\partial_\theta (\sin\theta\,  f^R)  +  \frac{1}{\sin \theta} \partial_\psi f^I & = 2 \pi \sqrt{2} \Delta\mathcal{J} \label{socutereal} \\
- a \sin \theta\,  \partial_v f^R +  \frac{1}{\sin \theta}\partial_\theta (\sin\theta\,  f^I) - \frac{1}{\sin \theta} \partial_\psi f^R &= 0. \label{socuteimag}
\end{align}
This form shows that solutions may be generated by taking $f^R$ to be any real function of $(v,\theta,\psi)$, using equation \eqref{socuteimag} to determine $f^I$, and letting equation \eqref{socutereal} define the current.   Putting everything together, the solution is
\begin{align}
\phi_0 & = \frac{f^R+if^I}{\Delta \rho} \label{phi0soln} \\
\mathcal{J} & = \frac{1}{2\pi \sqrt{2} \Delta}\left( a \sin \theta\,  \partial_v f^I + \frac{1}{\sin \theta}\partial_\theta (\sin\theta\,  f^R)  +  \frac{1}{\sin \theta} \partial_\psi f^I\right), \label{Jsoln}
\end{align}
where $f^R$ is a real function of $(v,\theta,\phi)$ and $f^I$ is related by
\begin{equation}
f^I = \frac{1}{\sin\theta}\int \left( a \sin^2 \theta\,  \partial_v f^R + \partial_\psi f^R \right) d \theta, \label{fIsoln}
\end{equation}
where we leave the integral indefinite.  While any function $f^R$ will give rise locally to a solution, this solution may become singular at the poles $\theta=\{0,\pi\}$, which are not covered by our coordinates. 

It is not difficult to find regular solutions by judicious choice of $f^R$.  For illustration,  a non-stationary, non-axisymmetric example is given by $f^R=15 F(v) \sin^2 \theta \cos \theta \cos \psi$, for a real function $F(v)$, which gives rise to the solution
\begin{equation}\label{prehistoric-beast}
\phi_0 = \frac{\sin^2 \theta}{\Delta \rho} \left\{ 5 F(v)(3 \cos \theta \cos \psi - i \sin \psi) + 3 i a \cos \psi \sin^2 \theta  F'(v) \right\}.
\end{equation}
The associated current, not displayed here, is regular at the poles.  Equation \eqref{prehistoric-beast}, together with $\phi_1=\phi_2=0$, gives a concrete example of a non-stationary, non-axisymmetric, globally regular solution to force-free electrodynamics in the Kerr background.  This solution represents a configuration of fields and plasma that propagates, unscattered, from past null infinity into the black hole.

A second approach to solving equation \eqref{socute} in the general
case is to make use of the symmetries of the underlying Kerr spacetime,
by expanding the field quantities in modes with harmonic $v$ and $\psi$ 
dependence.  For a single such mode,
\begin{align}
f(v,\theta,\psi) & = f_{m \omega}(\theta) e^{i m \psi} e^{-i \omega v}, \\
\mathcal{J}(v,r,\theta,\psi) & = \mathcal{J}_{m\omega}(r,\theta) e^{i m \psi} e^{-i \omega v},
\end{align}
equation \eqref{socute} becomes
\begin{equation}
\left( \partial_\theta +\cot\theta- \omega a \sin \theta + m\csc \theta \right) f_{m \omega}(\theta) = 2 \pi \sqrt{2} \Delta  \mathcal{J}_{m \omega}. \label{socuteinfourierspace}
\end{equation}
When the current vanishes the solution to \eqref{socuteinfourierspace} is $e^{- a \omega \cos \theta} \cot^m(\tfrac{\theta}{2})\csc\theta$.  This is singular at $\theta =0$ if $m\ge0$, and at $\theta=\pi$ if $m\le0$.  Hence, as in the Schwarzschild case, the existence of \textit{vacuum} solutions of this null, nonscattering type on the Kerr background is precluded by the requirement of global regularity. This conclusion also follows from the general analysis of Teukolsky and Press \cite{teukolsky-press1974}, which used spheroidal harmonic expansions for the NP scalars \cite{teukolsky-press1974}.  In particular, their analysis shows that $\phi_2=0$ implies $\phi_0=0$ (and vice versa) for regular fields.

For force-free solutions we only require that $\mathcal{J}$ is real, $\mathcal{J}_{m \omega}=\bar{\mathcal{J}}_{-m,-\omega}$. Equation \eqref{socuteinfourierspace} then yields
\begin{equation}
\left( \partial_\theta +\cot\theta- \omega a \sin \theta + m\csc \theta \right) f_{m \omega } =\left( \partial_\theta +\cot\theta+ \omega a \sin \theta - m\csc \theta \right) \bar{f}_{-m, -\omega},
\end{equation}
or equivalently,
\begin{equation}\label{ABeqn}
\left( \partial_\theta +\cot\theta\right)A_{m \omega}
= \left( \omega a \sin \theta - m\csc \theta \right)B_{m \omega},
\end{equation}
with $A_{m \omega}=(f_{m \omega}-\bar{f}_{-m, -\omega})$ and $B_{m \omega}=( f_{m \omega}+ \bar{f}_{-m,-\omega})$, which we define only for $m \geq 0$ and $\omega \geq 0$.  We may thus choose any complex function for each $A_{m \omega}$, and define $B_{m \omega}$ so that \eqref{ABeqn} is satisfied, which yields the solution
\begin{align}
f_{m\omega} & = \frac{1}{2} \left( A_{m \omega} + \frac{\sin\theta\, \partial_\theta A_{m \omega} + \cos\theta\, A_{m \omega}}{\omega a\sin^2\theta -m} \right) \label{fmw} \\
f_{-m,-\omega} & = -\frac{1}{2} \left( \bar{A}_{m \omega} - \frac{\sin\theta\, \partial_\theta \bar{A}_{m \omega} + \cos\theta\, \bar{A}_{m \omega}}{\omega a\sin^2\theta -m} \right). \label{fmwminus}
\end{align}
(For the $m=\omega=0$ mode, $B$ is freely chosen, while $A$ must vanish by regularity.) We thus obtain the general solution, in a mode expansion, in terms of free complex functions $A_{m \omega}$ with $m \geq 0$ and $\omega \geq 0$.  This solution is not always regular; however, for example, if $m>\omega a$ and $A$ and its derivative are regular, then regularity of $f$ (and $\phi_0$) is guaranteed.

If instead we regard the current as given, then the the general solution to equation \eqref{socuteinfourierspace} is obtained by the ``method of integrating factors'' using the homogeneous solution,
\begin{equation}
f_{\omega m} = 2\pi \sqrt{2} e^{- a \omega \cos \theta} \cot^m(\tfrac{\theta}{2}) 
\csc\theta\int (\Delta \mathcal{J}_{m \omega})e^{a \omega \cos \theta} \tan^m(\tfrac{\theta}{2}) \sin \theta d \theta, \label{fgen}
\end{equation}
where we leave the integral indefinite.  We have written $\Delta$ together with $\mathcal{J}_{m \omega}$ to remind the reader that the combination $\Delta \mathcal{J}_{m \omega}$ must be independent of $r$.  The solution for a general, conserved current may be built up from these modes in the usual way.  However, for some currents the associated $\phi_0$ will be singular at the poles.  

Finally, note that if the current is specified then one may instead use the Teukolsky equation \cite{teukolsky1973}, a second-order differential equation for $\phi_0$ (or alternatively $\phi_2$) with a source term constructed from the current and its first derivatives.  Expanding the source in spin-weighted spherodial harmonics, the Teukolsky radial equation may then be solved for each mode.

To summarize, we have given three representations of the general (non-symmetric) solution in Kerr.  Equations \eqref{phi0soln}-\eqref{fIsoln} give the solution in terms of a free real function $f^R(v,\theta,\psi)$; equations \eqref{fmw}-\eqref{fmwminus} give the mode functions in terms of free complex functions $A_{m \omega}(\theta)$; and equation \eqref{fgen} gives the mode functions in terms of a specified current.  While regularity at the poles is easy to achieve in specific cases, we have not determined general necessary and sufficient conditions for regularity.  By contrast, in the Schwarzschild case we have given a simple representation of the general regular solution in terms of a choice of a regular real function $S(v,\theta,\phi)$ (equation \eqref{phischwarz}), as well as a necessary and sufficient condition on the current for regularity of the solution (discussion below \eqref{phischwarz}).

\section{Outgoing Solutions}\label{sec:outgoing}

While it would be straightforward to repeat the calculations of the previous section for the outgoing congruence $\ell^\mu$, using outgoing coordinates instead of ingoing ones, it is more efficient and instructive to instead exploit the time reversal symmetry of the Kerr spacetime, which sends ingoing to outgoing.  In Boyer-Lindquist coordinates this diffeomorphism is defined by $t\rightarrow-t,\phi \rightarrow -\phi$, but it extends globally to the spacetime.  Denoting it by $\iota$, we have $\iota^*g_{\mu \nu}=g_{\mu \nu}$, expressing the symmetry of the Kerr spacetime.  Since the force-free equations are constructed covariantly out of the metric and field strength, to any solution $F_{\mu \nu}$ with current $J^\mu$ there exists a corresponding solution $\iota^* F_{\mu \nu}$ with current $\iota_* J^\mu$.  Since the diffeomorphism exchanges ingoing with outgoing, we thereby associate an outgoing solution with each ingoing one (and vice-versa).\footnote{Menon and Dermer demonstrated this correspondence for stationary, axisymmetric solutions by explicit calculation in a 3+1 framework \cite{menon-dermer2011}.}

The NP scalars \eqref{phi0}-\eqref{phi2} for the time reversed solution, using the original tetrad, are defined by 
expressions of the form $(\iota^*F_{\mu\nu})e_1^\mu e_2^\nu$, where $e_{1,2}^\mu$ are tetrad vectors.
Since $\iota\iota=1$, this is equivalent to $\iota^*(F_{\mu\nu}(\iota_*e_1^\mu)(\iota_*e_2^\nu))$.
For the Kinnersley tetrad, \eqref{ellbl}-\eqref{mbl}, we have 
\beq
\iota_*\ell^\mu=-\frac{2\Sigma}{\Delta} n^\mu,\quad 
\iota_*n^\mu=-\frac{\Delta}{2\Sigma} \ell^\mu,\quad 
\iota_*m^\mu = \frac{1}{\rho^2\Sigma}\bar m^\mu,\quad
\iota_*\bar m^\mu = {\rho^2\Sigma}\, m^\mu.
\eeq
Using these relations, together with the definitions of the NP scalars \eqref{phi0}-\eqref{phi2}, we find
\begin{align}
\phi_0^{\textrm{new}} & =  \frac{2}{\rho^2\Delta} \ \iota^*\phi_2 \label{phi0new} \\
\phi_1^{\textrm{new}} & = -\iota^*\phi_1 \label{phi1new} \\
\phi_2^{\textrm{new}} & = \frac{\rho^2\Delta }{2} \iota^*\phi_0. \label{phi2new}
\end{align}
Thus, at the level of Kinnersley tetrad NP scalars, one obtains 
the time-reversed solution by applying $\iota$ to the scalars 
(i.e., sending $t \rightarrow -t$ and $\phi \rightarrow -\phi$) and 
then interchanging $\phi_0$ and $\phi_2$ with the multiplying factors given above.  
In particular, for our ingoing solutions, where $\phi_2$ vanishes, 
this operation yields outgoing solutions, where $\phi_0$ vanishes.

In outgoing coordinates $(u,r,\theta,\chi)$ our outgoing solutions have 
$\phi_2 \propto r^{-1}$ and $\phi_1 \propto r^{-2}$ for large $r$.  The stress tensor then approaches the form $2\pi T^{EM}_{\mu \nu} = |\phi_2|^2 \ell_\mu \ell_\nu$, representing an outgoing energy flux.  In particular, the flux of Killing energy through an $r \rightarrow \infty$ sphere is given by\footnote{This limit is at fixed $u$, approaching future null infinity.  In the stationary case flux integrals are often computed at spatial infinity, $r \rightarrow \infty$ fixing $t$.  This is equivalent to our \eqref{Efluxout} for stationary solutions.}
\beq\label{Efluxout}
\lim_{r \rightarrow \infty} \int (T^{EM})^r{}_u\,  r^2 d \Omega 
=\frac{1}{8\pi} \lim_{r \rightarrow \infty} \int |2\phi_2|^2 r^2 d\Omega > 0,
\eeq
where $d\Omega$ is the two-sphere area element.  Whereas the ingoing solutions represent waves propagating in from past null infinity into the black hole, the outgoing solutions represent waves emerging from the past horizon and propagating to future null infinity.

As noted below equation \eqref{solved1}, the stationary ingoing solutions are all singular at the past horizon.  Since the diffeo interchanges the past and future horizon, the corresponding outgoing solutions (that is, all stationary outgoing solutions) will thus be irregular on the future horizon.  On the other hand, in the time-dependent case one may construct ingoing and outgoing solutions regular on both horizons by appropriate choice of the free functions appearing in our solutions.

\section{Rotating Magnetospheres}\label{sec:magnetospheres}

When the monopole charge is non-zero, $q\neq0$, the solutions are magnetically dominated, so that there exist frames where the electric field vanishes.  The Michel monopole and its generalizations are constructed from the outgoing solutions by imposing conducting boundary conditions (electric field vanishes in the frame of the conductor) at the surface of a star of radius $R$, inside of which one does not specify the fields.  For a conducting star rotating in the $\phi$ direction with constant angular velocity $\omega$, it is easy to check that the potential function must satisfy $S(\theta) = \sqrt{2} q \omega \cos \theta$, and this reproduces the original Michel solution \cite{michel1973b}.  In particular, choosing $q = B_0 R^2$ in equation \eqref{phi1soln} and $S(\theta) = \sqrt{2} B_0 R^2 \omega \cos \theta$ in equation \eqref{phischwarz} and applying time reversal yields Michel's solution in the notation of \cite{beskin-book} (see equation (2.228) therein).  Here $B_0$ is the (constant) strength of the magnetic monopole field at the surface of the star.  Notice that the null solution $q=0$, corresponding to the outgoing Menon/Dermer solution \cite{menon-dermer2011} in flat spacetime, is obtained only as a singular limit of this choice of parameters: $B_0 R^2 \rightarrow 0$ and  $\omega \rightarrow \infty$, with $B_0 R^2 \omega$ held fixed.

The general choice of stationary, axisymmetric potential function $S(\theta)$ corresponds to a star undergoing differential ($\theta$-dependent) rotation, and was found by by Blandford and Znajek \cite{blandford-znajek1977} in flat spacetime in the outgoing case (see equation (6.4) therein).  Blandford and Znajek also comment that they have found a generalization to Schwarzschild, which is not regular on the event horizon (see section 5 therein).  They presumably refer to the general choice of stationary, axisymmetric potential function $S(\theta)$ in Schwarzschild spacetime in the outgoing case, which indeed produces a solution singular on the horizon.  Lyutikov \cite{lyutikov2011} wrote out this solution and attributed it to Blandford and Znajek, considering it as the exterior solution of a star.  He also found the generalization allowing the function $S$ to depend on time, corresponding to a time-dependent angular velocity of the star.  In this case $S$ becomes a free function of outgoing Eddington-Finklestein time, i.e., $S=S(u,\theta)$. 

Our most general solution allows the free function to depend on $\phi$ as well, $S(u,\theta,\phi)$ (and extends to the Kerr background for the case of zero monopole charge $q$).  By finding frames in which the electric field vanishes (which is possible only with $q\neq 0$) we may match these solutions to conducting stars undergoing more complicated rotational motions than the axisymmetric rotation present in the Michel and Lyutikov solutions.  In fact, we may allow the axis and speed of rotation to vary arbitrarily, corresponding to the most general rigid motion of a conducting star whose origin is fixed.  This construction, together with the dynamics of the current sheet entailed by splitting the monopole, will be presented in a future publication.

\ack{We wish to thank Jonathan McKinney for helpful conversations. S.G. acknowledges support from NASA through the Einstein Fellowship Program, Grant PF1-120082. D.B. and T.J. were supported in part by the NSF under grant No. PHY-0903572.}

\appendix

\section{The Kinnersley Tetrad in Flat Spacetime.}

In this appendix we work in flat spacetime and relate the NP quantities used in the paper to ordinary electric and magnetic fields $\vec{E}$ and $\vec{B}$.  We decompose these vectors in the orthonormal frame $\{\hat{r},\hat{\theta},\hat{\phi}\}$ associated with spherical coordinates as $\vec{E} = E_{\hat{r}} \hat{r} + E_{\hat{\theta}} \hat{\theta} + E_{\hat{\phi}} \hat{\phi}$ and $\vec{B} = B_{\hat{r}} \hat{r} + B_{\hat{\theta}} \hat{\theta} + B_{\hat{\phi}} \hat{\phi}$.  The relationship to the spherical-coordinate components of the field strength $F_{\mu \nu}$ is given by
\begin{align}
E_{\hat{r}} & = -F_{tr} & B_{\hat{r}} & = \frac{1}{r^2 \sin \theta} F_{\theta \phi} \label{EB1} \\
E_{\hat{\theta}} & = \frac{-1}{r} F_{t\theta} & B_{\hat{\theta}} & = \frac{-1}{r \sin \theta} F_{r \phi} \\
E_{\hat{\phi}} & = \frac{-1}{r \sin \theta} F_{t\phi} & B_{\hat{\phi}} & = \frac{1}{r} F_{r \theta}. \label{EB3}
\end{align}
Specialized to flat spacetime, and in spherical coordinates, the Kinnersley tetrad \eqref{ellbl}-\eqref{mbl} becomes
\begin{align}
\ell^\mu & = \left( 1, 1, 0, 0 \right) \label{ellflat} \\
n^\mu & = \left( \tfrac{1}{2}, -\tfrac{1}{2}, 0, 0 \right) \label{nflat} \\
m^\mu & = \frac{1}{\sqrt{2}r} \left( 0, 0, 1, \frac{i}{\sin \theta} \right), \label{mflat}
\end{align}
so that, by equations \eqref{phi0}-\eqref{phi2}, the NP scalars are
\begin{align}
\phi_0 & = \frac{1}{\sqrt{2}}\left( -E_{\hat{\theta}} + B_{\hat{\phi}} - i ( E_{\hat{\phi}} + B_{\hat{\theta}} ) \right) \label{phi0flat} \\
\phi_1 & = \frac{1}{2} \left( E_{\hat{r}} + i B_{\hat{r}} \right) \label{phi1flat} \\
\phi_2 & = \frac{1}{2\sqrt{2}}\left( E_{\hat{\theta}} + B_{\hat{\phi}} + i ( -E_{\hat{\phi}} + B_{\hat{\theta}} ) \right). \label{phi2flat}
\end{align}
A spherical wave has $\vec{E}$ and $\vec{B}$ orthogonal, equal in magnitude, and  tangent to the sphere.  For an outgoing wave $\phi_0$ will vanish, while real and imaginary parts of $\phi_2$ will represent the two polarizations of the wave.  Similarly, for an ingoing wave $\phi_2$ will vanish, while the real and imaginary parts of $\phi_0$ will give the two polarizations.  In this sense $\phi_0$ describes to ingoing radiation, while $\phi_2$ describes outgoing radiation.  The remaining scalar $\phi_1$ describes radial fields.  In the asymptotic (large $r$) region of Kerr the Kinnersley tetrad goes over to its flat spacetime form, and these properties are preserved: $\phi_0$ and $\phi_2$ are adapted to radiation entering and leaving the spacetime, respectively.

\section*{References}

\bibliographystyle{iopart-num}

\bibliography{nullff}{}

\end{document}